\documentclass{article}
\usepackage{spconf,amsmath,graphicx,hyperref}
\usepackage{amssymb}
\usepackage{algorithm,algorithmic}
\usepackage{adjustbox}
\usepackage{makecell} 

\title{MK-SGC-SC: Multiple Kernel Guided Sparse Graph Construction in Spectral Clustering for Unsupervised Speaker Diarization}

\name{Nikhil Raghav$^{1,2}$, Avisek Gupta$^1$, Swagatam Das$^{1,3}$, Md Sahidullah$^1$}
\address{$^1$Institute for Advancing Intelligence, TCG CREST, Kolkata -- 700091, India\\
$^2$Department of Computer Science, RKMVERI, Howrah -- 711 202, India\\
$^3$Electronics and Communication Sciences Unit, Indian Statistical Institute, Kolkata-700 108, India}

\begin{document}
\ninept
\maketitle
\begin{abstract}
Speaker diarization aims to segment audio recordings into regions corresponding to individual speakers. Although unsupervised speaker diarization is inherently challenging, the prospect of identifying speaker regions without pretraining or weak supervision motivates research on clustering techniques. In this work, we share the notable observation that measuring multiple kernel similarities of speaker embeddings to thereafter craft a sparse graph for spectral clustering in a principled manner is sufficient to achieve state-of-the-art performances in a fully unsupervised setting. Specifically, we consider four polynomial kernels and a degree one arccosine kernel to measure similarities in speaker embeddings, using which sparse graphs are constructed in a principled manner to emphasize local similarities. Experiments show the proposed approach excels in unsupervised speaker diarization over a variety of challenging environments in the DIHARD-III, AMI, and VoxConverse corpora. To encourage further research, our implementations are available at \url{https://github.com/nikhilraghav29/MK-SGC-SC}.
\end{abstract}
\vspace{-0.2cm}
\section{Introduction} \label{sec:intro}
\vspace{-0.2cm}
Accurately identifying and attributing segments of speech to individual speakers within an audio stream is a crucial component of modern-day audio analytics, especially in scenarios involving multi-speaker conversational speech. A classical unsupervised machine learning problem, formally known as \emph{speaker diarization} (SD)~\cite{anguera2012speaker, park2022review}, also popularly known as ``\emph{who spoke when}", addresses this challenge. It is an enabling technology for numerous downstream applications, such as speaker-attributed meeting transcription, doctor-patient interactions in healthcare, call center analytics, courtroom proceedings, forensics, etc. Furthermore, its potential is exploited in many closely-related speech processing applications, for instance, automatic speech recognition (ASR)~\cite{kanda19_asru}, and speaker separation~\cite{raj21_slt}. 
A modular pipeline for diarization conventionally consists of modules spanning across several stages~\cite{GMM}, namely \emph{speech enhancement}, \emph{speech activity detection} (SAD) to identify the speech regions, \emph{segmentation}, \emph{speaker embedding extraction}, \emph{clustering}, and \emph{re-segmentation}.

The use of neural networks (NNs) has been explored in the context of diarization, often focusing on individual modules in the conventional diarization pipeline~\cite{Yella2014speaker, snyder2018x, deep_neural_embeddings}. End-to-end neural diarization methods (EEND)~\cite{yang2024icassp, fujita2020end, Fujita2019EndtoEndNS, Fujita2019EndtoEnd, horiguchi2022encoder} have also been investigated as an integrated system to perform diarization. 
 
However, end-to-end approaches face key limitations. They demand large amounts of annotated data, which is scarce for multi-speaker conversations. Moreover, they struggle to generalize across scenarios (e.g., telephone vs. restaurant recordings) and to settings with an unknown number of speakers. An attractive alternative to EEND models is unsupervised diarization, where spectral clustering~\cite{von2007tutorial} is perhaps the most widely used approach~\cite{Park2020AutoTuningSC,dawalatabad21_interspeech}. Spectral clustering offers key advantages: as an unsupervised method, it requires no extra training data, avoids issues of overfitting and generalizability, and integrates easily into the SD pipeline.

Previous works on spectral clustering approaches for SD of $n$ speech segments operate by first forming $n^2$-sized similarity matrices. The entries of these matrices are significantly pruned to thereby form sparse similarity matrices. In spectral clustering, pruning removes noisy edges from the similarity graph, yielding sparser affinity matrices from which clusters of same-speaker regions can be identified more reliably~\cite{Park2020AutoTuningSC}. This observation motivated semi-supervised spectral clustering (SS-SC)~\cite{dawalatabad21_interspeech}, where a separate set of \emph{development} set of speech segments are consulted to determine the degree of pruning that should be performed. The performance of the method is then tested on the \emph{evaluation} set. A notable unsupervised SD approach is auto-tuning spectral clustering (ASC)~\cite{Park2020AutoTuningSC}, which analyzes multiple prunings of the similarity matrix to automatically determine the optimal pruning level. However, ASC incurs high computational cost. To address this, the recent \emph{spectral clustering on p-neighborhood retained affinity matrix} (SC-pNA)~\cite{SC-pNA} introduced an efficient $k$-means–based approach to prune each row of the similarity matrix.

In this work, we propose an efficient yet effective alternative to existing similarity matrix pruning approaches. Our approach involves constructing multiple suitable $n^2$-sized similarity matrices, whose entries are computed from kernel functions. From each of these similarity matrices, a principled approach is employed to construct a fused sparse graph adjacency matrix. The process of sparsification involves scaling similarity magnitudes to similar scales, as well as consulting neighborhoods of each speech segment embedding. The sparse adjacency matrix is then provided to a standard spectral clustering approach, which estimates the number of clusters and finally identifies the clusters. Extensive experiments show that our proposed \emph{multiple kernel guided sparse graph construction in spectral clustering} (MK-SGC-SC) is highly effective for SD, consistently outperforming unsupervised methods and often matching or surpassing semi-supervised ones.

Our main contributions are: (i)~introducing multiple kernel similarities tailored for SD, (ii)~proposing a principled framework to build, sparsify, and fuse graph adjacency matrices from these kernels, and (iii)~demonstrating through experiments that MK-SGC-SC consistently outperforms state-of-the-art unsupervised methods and often matches semi-supervised SD.

\vspace{-0.2cm}
\section{Related Works}
\vspace{-0.2cm}
\label{sec:related_works}
In speaker diarization, several clustering algorithms such as \emph{k-means}, \emph{agglomerative hierarchical clustering}, \emph{mean-shift}~\cite{salmun2017plda}, and \emph{spectral clustering} (SC)~\cite{ning2006spectral} have been explored, with SC emerging as the most widely used, particularly in modern systems leveraging deep speaker embeddings~\cite{Park2020AutoTuningSC}, due to its simplicity, low parameterization, and strong theoretical basis in spectral graph theory~\cite{ng2001spectral,von2007tutorial}. Another line of research investigates supervised~\cite{fully_supervised} and self-supervised approaches~\cite{SSL_SD,ssl}, including graph-based hierarchical methods such as \emph{path integral clustering} (PIC)~\cite{path_integral}. These methods, however, are inherently data-driven and rely strongly on the characteristics of the training set.  

Beyond conventional clustering algorithms, multiple kernel clustering (MKC) has shown considerable success \cite{oplfmvc,li2022local}, where optimization-based methods exploit information from multiple kernel similarities and fuse them, along with problem-specific cues, to obtain a consensus clustering \cite{liu2017optimal}. Furthermore, since data relationships can naturally be represented as graph edges, there has been significant interest in graph sparsification methods aimed at efficiently reducing graph complexity while preserving structural properties, including both heuristic approaches and algorithms with provable guarantees \cite{zhang2024digrass,hashemi2024comprehensive,sachdeva2023better}. This work advances spectral clustering by incorporating multiple kernel clustering and graph sparsification.

\vspace{-0.2cm}
\section{Proposed Methodology}
\label{sec:methodology}
\vspace{-0.2cm}

The proposed MK-SGC-SC operates in three stages: (i) \textbf{compute multiple kernel similarities} on speaker embeddings; (ii) \textbf{construct a sparse graph} by treating these kernels as adjacency matrices, sparsifying them via $k$-nearest neighborhoods, and fusing them into a single adjacency matrix; and (iii) \textbf{apply spectral clustering} to estimate the clusters.

A kernel matrix is a symmetric positive semi-definite (PSD) matrix $K \in \mathbb{R}^{n\times n}$, where each entry is $K_{ij}=f(x_i,x_j)$ for some PSD kernel function $f$. Kernel functions are often parametric, examples include polynomial kernels $f_{a,b}(x_i,x_j) = (x_i^Tx_j + a)^b$, arccosine kernels $f(x_i,x_j) = \frac{1}{\pi}\|x_i\|_2\|x_j\|_2(\sin \theta_{ij} + (\pi - \theta_{ij})\cos \theta_{ij})$ where $\theta_{ij} = \arccos((x_i^Tx_j)/(\|x_i\|_2\|x_j\|_2))$, etc. Kernel functions are interpreted as similarity measures between data instances; a suitable kernel best represents the similarities for the problem at hand.

Next, for a graph $G=(V,E)$ consisting of vertices $V$ and edges $E$, each entry of a graph adjacency matrix $A_{ij} \in \mathbb{R}$ represents the weight of an edge between nodes $v_i,v_j \in V$. In general, $A$ does not need to be PSD; for undirected graphs, $A$ is symmetric. While an adjacency matrix need not be a kernel matrix, any kernel matrix can be a graph adjacency matrix. A complete graph has all $\binom{n}{2}$ edges between its vertices, in comparison, a sparse graph has significantly fewer edges, thus a sparse adjacency matrix describing a sparse graph has most of entries as zero. A challenging question in machine learning is on how a general graph should be suitably sparsified to form a sparse graph~\cite{hashemi2024comprehensive}.

Finally, the spectral clustering method operates on graph adjacency matrices. While kernel matrices can be considered, usually clusters are easier to identify from sparse graph adjacency matrices, therefore forming sparse $A$ is generally encouraged, which can be observed even in the literature on spectral clustering for spearker diarization. However, kernel matrices are usually not sparse. 

With this background, we describe the steps of the MK-SGC-SC method next.

\textbf{Step 1.} Form $m$ symmetric and PSD kernel similarity matrices $\{K^1,...,K^m\}$, each representing a distinct kernel similarity. Here kernels should be selected that are suitable for the problem. For diarization, the measure should help distinguish between embeddings of the same speaker in comparison to other speakers. Identifying a suitable measure is a challenge, an ideal similarity measure should generalize across embeddings of speakers and challenging recording environments.

\textbf{Step 2.} From the kernels $\{K^l\}_{l=1}^{m}$, initial graph adjacency matrices $\{A^l\}_{l=1}^{m}$ are created in the following manner. First, the kernel entries are shifted and scaled to obtain $A^l_{ij} = (K^l_{ij} - \min_{ij}\{K^l_{ij}\})/||K^l||_F$. Shifting makes all $A^l_{ij} \geq 0$, while scaling ensures that one adjacency matrix does not have larger magnitude values compared to others. This is relevant during the fusion of matrices, where larger magnitudes of one matrix can dominate over others, making the smaller magnitude matrices irrelevant.

\textbf{Step 3.} Remove self-loops: Set $A^l_{ii}=0\ \forall i,l$. The similarity of a vertex with itself can be of much larger magnitude compared to the similarities towards others; removing these similarities lets other similarities contribute better.

\textbf{Step 4.} Sparsify by consulting k-nearest neighbors: $A^l_{ij}=0$ $\forall x_j \notin \mathcal{N}_k(x_i)$. As $A^l$ contains values derived from kernel similarities, in general it will not be sparse. This step sparsifies $A^l$ to reinforce local connections between vertices; the majority of weak connections between vertices are removed.

\textbf{Step 5.} Fuse adjacency matrices $A^*_{ij} = \frac{1}{m} \sum_{l=1}^{m}A^l_{ij}$. If $A^l$ are suitable sparse representations of inter-speaker similarities, then this equal-weighted fusion creates a consensus sparse affinity matrix.

\textbf{Step 6.} Scale the fused matrix: $A^*_{ij} = A^*_{ij} / ||A^*||_F\ $. This scaling ensures numerical stability of the eigen-decomposition of the Laplacian of A.

The following steps are similar to standard spectral clustering, where the number of clusters are estimated from the eigengap~\cite{chung97}.

\textbf{Step 7.} Form the unnormalized Laplacian: $L = D - A^{*}$, where $D_{ii} = \sum_{j=1}^{n} A^*_{ij}$, $D_{ij} = 0$ for $i\neq j$.

\textbf{Step 8.} Consider the eigenvalues of $L$ in ascending order: $0=\lambda_1\leq...\leq\lambda_n$. We form the $(M$-$1)$-length eigengap vector as $e_{\text{gap}}=[\lambda_2-\lambda_1,\lambda_3-\lambda_2,...,\lambda_{k+1}-\lambda_{k}...,\lambda_M-\lambda_{M-1}]$, where $M=\min\{k_{\max},n\}$. Here $k_{\max}$ can be specified by the user. We estimate the number of clusters as the maximum eigengap: $k^*=i$, if $(\lambda_{i+1}-\lambda_{i}) \geq (\lambda_{j+1}-\lambda_{j})\ \forall j \neq i$.

\textbf{Step 9.} Let $H \in\mathbb{R}^{n\times k^*}, H^TH=I_{k^*}$, contain the eigenvectors of L corresponding to the $k^*$ smallest eigenvalues. On the rows of $H$, we run $k$-Means clustering to identify $k^*$ number of clusters.

The complete algorithm of MK-SGC-SC is described below in Algorithm \ref{alg:mk-sc}.

\begin{algorithm}[!htb]
\footnotesize
\caption{MK-SGC-SC}
\label{alg:mk-sc}
\begin{algorithmic}[1]
\STATE Input: Speaker embeddings $\{e_p \in \mathbb{R}^{d}\}_{p=1}^{n}$, no. of neighbors $c$, $k_{\max}$.
\STATE Output: Cluster labels $\mathcal{C} \in \{1,\dots,k\}^n$
\FOR{$i,j = 1, \dots, n$}
    \STATE $K^1_{ij} := (e_i^Te_j)^2$, $K^2_{ij} := (e_i^Te_j + 1)^2$, $K^3_{ij} := (e_i^Te_j)^3$,
    \STATE $K^4_{ij} := (e_i^Te_j + 1)^3$, 
    \STATE $K^5_{ij} := \frac{1}{\pi}\|e_i\|_2\|e_j\|_2(\sin \theta_{ij} + (\pi - \theta_{ij})\cos \theta_{ij})$,
    \STATE where $\theta_{ij} = \arccos\left(\frac{e_i^Te_j}{\|e_i\|_2\|e_j\|_2}\right)$.
\ENDFOR
\FOR{$l = 1, \dots, m$}
    \STATE $A^l_{ij} := (K^l_{ij} - \min_{ij}\{K^l_{ij}\})/||K^l||_F$.
    \STATE $A^l_{ii} := 0\ \forall i$, \hspace{1.5em} $A^l_{ij}:=0$, $\forall x_j \notin \mathcal{N}^l_c(x_i)$. 
\ENDFOR
\STATE $A^*_{ij} := \frac{1}{m}\sum_{l=1}^m A^l_{ij}$, \hspace{1.5em} $A^*_{ij} := A^*_{ij} / \|A^*\|_F$.
\STATE // \textbf{Spectral Clustering}
\STATE $L := D - A^*$, where $D = \mathrm{diag}(A^*\mathbf{1})$.
\STATE Compute the eigenpairs $(\lambda_i, u_i)$ of $L$, sorted by $\lambda_1 \le \cdots \le \lambda_n$; estimate $k^*$ as the maximimum eigengap from $\lambda_1,\dots,\lambda_{M}$, where $M=\min\{n,k_{\max}\}$.
\STATE Form $H$ := $\left(u_1,...,u_{k^*}\right)$; apply $k$-Means on $H$ to obtain $\mathcal{C}$.
\end{algorithmic}
\end{algorithm}

\textbf{Computational Complexity}: Computing each kernel similarity of the embeddings requires $O(n^2d)$ time and $O(n^2)$ space. In general for m kernels, $O(mnd^2)$ time and $O(mn^2)$ space is required. 
%Here $m=5$ incurs $O(5nd^2)$ time and $O(5n^2)$ space complexity. 
Next for $A^l$, computations such as $\|K^l\|_F$ and $
min\{K^l_{ij}\}$ require at most $O(n^2)$ time. Computing $\mathcal{N}^l_c(x_i)$ takes $O(cn)$ time for every vertex, and thus needs a total of $O(cn^2)$ time. The eigen-decomposition of $L$ incurs $O(n^3)$ time and $O(n^2)$ space. Finally, $T$ iterations of $k$-Means clustering takes only $O(k^2nT)$ time. As usually $k\leq k_{\max}\ll n$, the overall time complexity of MK-SGC-SC is $O(n^3)$, as is in standard spectral clustering~\cite{von2007tutorial}. MK-SGC-SC requires $O(mn^2)$ space.

\vspace{-0.25cm}
\section{Experimental Setup}
\label{sec:experiments}
\vspace{-0.25cm}

We evaluate all methods on conversational speech corpora: (i) DIHARD-III~\cite{ryant21_interspeech} contains speech utterances from eleven diverse domains, ranging from a controlled scenario like \emph{audiobooks} and \emph{broadcast news} to challenging scenarios like \emph{restaurant} and \emph{web-video}. (ii) AMI~\cite{AMI_meeting_corpus} for the official Full-corpus-ASR partition, with TNO meetings excluded, contains recordings for 3-5 speakers, averaging 20-60 minutes, sampled at 16kHz for both close-talking and far-field microphones. (iii) VoxConverse~\cite{chung2020spot}, from the 2023 VoxSRC challenge, consists of over 50 hours of nearly 450 speakers sourced from YouTube videos. Each corpus includes separate development and evaluation splits, both treated as independent test sets in the unsupervised diarization setting.

All experiments are built using a modified version of the AMI recipe provided in the SpeechBrain~\cite{speechbrain_v1, speechbrain} speech toolkit~\footnote{\url{https://github.com/speechbrain/speechbrain/tree/develop/recipes/AMI}}. It extracts speaker embeddings using an ECAPA-TDNN model~\cite{ecapa2020}, pre-trained on the VoxCeleb dataset~\cite{nagrani17_interspeech,chung18b_interspeech}. All speech segments are of length 3s, with an overlap of 1.5s. The extracted embeddings taken from the penultimate layer of the ECAPA-TDNN model are $192$ dimensional. Diarization performance is evaluated using diarization error rate (DER), evaluated against ground-truth annotations, conducted with a 0.25s collar of tolerance. 

\vspace{-0.2cm}
\section{Results and Discussion}
\label{sec:results}
\vspace{-0.2cm}

We first investigate the unsupervised diarization performance of MK-SGC-SC, comparing against the SOTA spectral clustering approaches of SC-$p$NA and ASC. Next, the unsupervised MK-SGC-SC is compared with a Semi-Supervised SC (SS-SC) approach. This is followed by a study on the diarization performances of different kernels, after which ablation studies are conducted, and the effect of the neighborhood parameter $c$ is studied.

\noindent \textbf{Unsupervised SC-based diarization:} Table \ref{tab:unsup_sota_igoverlap} records the measured DERs (excluding overlapped speech regions) achieved by the unsupervised approaches of SC-$p$NA, ASC, and the proposed MK-SGC-SC. From the results, we observe that for a majority of the data splits (14 out of 30), MK-SGC-SC achieves lower DER compared to SC-$p$NA and ASC. The reduction in DER achieved by MK-SGC-SC is often significant compared to others; examples of reductions between the `lowest / next-lowest' include: court (eval) 1.99 / 6.36, restaurant (dev) 20.97 / 23.95, Array1-01 (dev) 2.64 / 5.44, VoxConverse (dev) 2.83 / 7.25 and (eval) 4.27 / 9.38. We also note that MK-SGC-SC usually achieves close to the lowest DERs for the splits on which other methods achieve the lowest DER.

\begin{table}[!htb]
\centering
\vspace{-6pt}
\caption{Speaker diarization performance in terms of DER ($\downarrow$) (overlapped regions ignored) achieved by the unsupervised methods.}
\label{tab:unsup_sota_igoverlap}
\adjustbox{max width=\columnwidth}{
\begin{tabular}{ll|cc|cc|cc|}
\cline{3-8}
    & \multicolumn{1}{l}{}              & \multicolumn{2}{|c}{MK-SGC-SC}   & \multicolumn{2}{c}{SC-pNA}    & \multicolumn{2}{c|}{ASC} \\
    \cline{3-8}
    & \multicolumn{1}{l}{}              & \multicolumn{1}{|c}{Dev}            & Eval           & Dev           & Eval          & Dev        & Eval       \\
    \hline
    \multicolumn{1}{|l}{} & broadcast & \textbf{1.07}  & \textbf{1.95}  & 2.36          & 4.07          & 3.54         & 2.92          \\
    \multicolumn{1}{|l}{} & court     & 3.85  & 1.99  & 5.29          & 6.36          & \textbf{1.91}          &\textbf{1.94}          \\
    \multicolumn{1}{|l}{} & cts                               & 2.48           & \textbf{2.07}  & \textbf{2.44} & 2.08 & 3.90          & 2.80          \\
    \multicolumn{1}{|l}{} & maptask                           & \textbf{1.47}  & 0.76           & 2.33          & \textbf{0.74} & 4.39          & 7.34          \\
    \multicolumn{1}{|l}{} & meeting                           & \textbf{3.58}  & 9.35 & 4.29          & 11.65         & 5.81          & \textbf{8.86}           \\
\multicolumn{1}{|c}{DIHARD-III}  & socio-lab & 1.36            & 1.01           & \textbf{1.20}  & \textbf{0.99} & 2.28          & 3.16          \\
    \multicolumn{1}{|l}{} & webvideo                          & \textbf{13.59} & \textbf{21.13} & 14.30          & 22.83         & 25.66          & 27.83          \\
    \multicolumn{1}{|l}{} & restaurant                        & 20.97 & 22.81 & 23.95         & 24.88         & \textbf{14.79}          & \textbf{18.69}          \\
    \multicolumn{1}{|l}{} & audiobooks                        & 1.66           & 0.31           & \textbf{0.08} & \textbf{0.09} & 23.33         & 26.62          \\
    \multicolumn{1}{|l}{} & clinical                          & \textbf{3.41}  & 2.61           & 4.24          & \textbf{2.58} & 9.60         & 9.78          \\
    \multicolumn{1}{|l}{} & socio-field                       & 3.56  & 2.41           & 4.68          & \textbf{2.35} & \textbf{2.59}          & 8.93          \\
    \hline
    \multicolumn{1}{|l}{} & Mix-Headset                       & 1.90   & \textbf{1.67}  & \textbf{1.77}          & 1.82          & 3.63          & 3.22          \\
\multicolumn{1}{|c}{AMI}         & Mix-Lapel                         & 2.26           & \textbf{2.29}  & \textbf{2.24} & 2.36          & 2.81          & 2.60         \\
    \multicolumn{1}{|l}{} & Array1-01                         & \textbf{2.64}  & \textbf{3.47}  & 5.44          & 4.38          & 4.14         & 3.67          \\
    \hline
\multicolumn{1}{|c}{VoxConverse} & Overall                           & \textbf{2.83}  & \textbf{4.27}  & 7.25          & 9.38          & 9.43          & 7.16       \\
\hline
\end{tabular}
}
\end{table}

Compared to the previous, we now consider the more challenging setting where DER is evaluated over overlapped speech regions as well. Consequently, in Table \ref{tab:unsup_sota_overlap}, we generally observe increased DERs for all methods. ASC is omitted here due to its higher DER. Notably, in this more challenging setting, MK-SGC-SC clearly outperforms SC-pNA, achieving the lowest DERs for 22 out of the 30 data splits. Moreover, MK-SGC-SC clearly outperforms SC-pNA on VoxConverse for both Dev and Eval sets. The exceptional performances observed clearly demonstrate the merit of considering multiple kernel similarities to guide the principled graph construction procedure for spectral clustering. Subsequent experiments are all performed on this challenging setting of measuring DER over overlapped speech regions.

\begin{table}[!htb]
\centering
\vspace{-0.4cm}
\renewcommand{\arraystretch}{1.1}
\caption{Speaker diarization performance in terms of DER($\downarrow$) (overlapped regions included) achieved by the unsupervised methods.}
\label{tab:unsup_sota_overlap}
\adjustbox{max width=0.85\columnwidth}{
\begin{tabular}{ll|cc|cc|}
\cline{3-6}
& \multicolumn{1}{l}{}              & \multicolumn{2}{|c}{MK-SGC-SC}   & \multicolumn{2}{c|}{SC-pNA} \\
\cline{3-6}
& \multicolumn{1}{l}{}              & \multicolumn{1}{|l}{Dev} & Eval  & Dev & Eval \\
\hline
\multicolumn{1}{|l}{} & broadcast & \textbf{1.64}  & \textbf{2.70}   & 2.98 & 4.77 \\
\multicolumn{1}{|l}{} & court     & \textbf{4.64}  & \textbf{2.79}  & 6.04          & 7.15 \\
\multicolumn{1}{|l}{} & cts       & \textbf{8.20}   & \textbf{6.62}  & 8.22 & 6.63 \\
\multicolumn{1}{|l}{} & maptask   & \textbf{1.88}  & 0.94           & 2.73          & \textbf{0.92} \\
\multicolumn{1}{|l}{} & meeting   & \textbf{16.29} & \textbf{19.09} & 16.79 & 21.26 \\
\multicolumn{1}{|c}{DIHARD-III} & socio-lab & 3.23  & 2.00    & \textbf{3.08} & \textbf{1.97} \\
\multicolumn{1}{|l}{} & webvideo  & \textbf{29.77} & \textbf{31.51} & 30.68         & 33.06 \\
\multicolumn{1}{|l}{} & restaurant & \textbf{35.07} & \textbf{35.04} & 37.28         & 38.81 \\
\multicolumn{1}{|l}{} & audiobooks & 1.66  & 0.31           & \textbf{0.08} & \textbf{0.09} \\
\multicolumn{1}{|l}{} & clinical   & \textbf{5.51}  & 3.35  & 6.31          & \textbf{3.31} \\
\multicolumn{1}{|l}{} & socio-field& \textbf{8.22}  & 3.86           & 9.21          & \textbf{3.79} \\
\hline
\multicolumn{1}{|l}{} & Mix-Headset & 14.93 & \textbf{17.33} & \textbf{14.84} & 17.44 \\
\multicolumn{1}{|c}{AMI} & Mix-Lapel   & \textbf{15.23} & \textbf{17.87} & \textbf{15.23} & 17.91 \\
\multicolumn{1}{|l}{} & Array1-01   & \textbf{15.62} & \textbf{18.98} & 17.98 & 19.62 \\
\hline
\multicolumn{1}{|c}{VoxConverse} 
& Overall     & \textbf{5.12}  & \textbf{5.82}  & 9.41   & 10.83 \\
\hline
\end{tabular}
}
\end{table}

\textbf{Semi-supervised SC-based diarization:} We conduct a contrastive study of the unsupervised MK-SGC-SC, and the semi-supervised SS-SC, where the latter is tuned on a separate `Dev' set and evaluated on the `Eval' set. We compare both methods under the setting where the number of clusters ($k^*$) is estimated, as well as one where the method is provided $k^*$ by an oracle. Interestingly, for both cases MK-SGC-SC has comparable performance to SS-SC, on both the Dev set and the Eval set, and often even outperforms SS-SC. When $k^*$ is estimated, on DIHARD-III MK-SGC-SC outperforms SS-SC for six out of 11 Dev sets; barring restaurant, it provides comparable performances on the remaining. Similarly MK-SGC-SC excels on six of the 11 Eval sets, and generally achieves DER comparable to SS-SC. On AMI, SS-SC shows slightly lower DER. When the oracle $k^*$ is provided to both methods, we generally observe MK-SGC-SC either outperforming SS-SC, otherwise the difference in DER between the two are quite marginal. This leads us to an important conclusion: MK-SGC-SC, when operating in an unsupervised manner, is a viable alternative to the semi-supervised SS-SC, as it often outperforms SS-SC or achieves comparable performance.

\begin{table}[!htb]
\centering
\vspace{-0.4cm}
\renewcommand{\arraystretch}{1.1}
\caption{Comparing DER($\downarrow$) with semi-supervised SS-SC, where $k^*$ is either (i) estimated or (ii) oracle-provided.}
\label{tab:semisup_sota_overlap}
\adjustbox{max width=\columnwidth}{
\begin{tabular}{ll|cccc|cccc|}
\cline{3-10}
& \multicolumn{1}{l|}{}              & \multicolumn{4}{c|}{estimated $k^*$}   & \multicolumn{4}{c|}{oracle $k^*$} \\
& \multicolumn{1}{l|}{}              & \multicolumn{2}{c}{MK-SGC-SC}   & \multicolumn{2}{c|}{SS-SC}       & \multicolumn{2}{c}{MK-SGC-SC} & \multicolumn{2}{c|}{SS-SC} \\
\cline{3-10}
& \multicolumn{1}{l|}{}              & Dev            & Eval           & Dev            & Eval           & Dev              & Eval            & Dev            & Eval          \\
\hline
\multicolumn{1}{|l}{} & broadcast & \textbf{1.64}  & \textbf{2.70}   & 2.03  & 3.58           & 1.52             & \textbf{3.50}    & \textbf{1.40}   & 3.63          \\
\multicolumn{1}{|l}{} & court     & 4.64  & 2.79  & \textbf{2.01}  & \textbf{2.09}  & 1.86    & \textbf{1.69}   & \textbf{1.77}  & 2.18          \\
\multicolumn{1}{|l}{} & cts                               & \textbf{8.20}   & 6.62  & 8.28  & \textbf{6.58}  & \textbf{8.18}    & 6.62   & 8.22           & \textbf{6.57} \\
\multicolumn{1}{|l}{} & maptask                           & \textbf{1.88}  & \textbf{0.94}  & 2.19           & 1.78  & \textbf{1.85}    & \textbf{0.94}   & 3.45           & 2.60           \\
\multicolumn{1}{|l}{} & meeting                           & 16.29 & 19.09 & \textbf{16.11} & \textbf{16.79} & \textbf{16.19}   & \textbf{15.33}  & 16.65          & 15.88         \\
\multicolumn{1}{|c}{\small DIHARD-III}  & socio-lab & \textbf{3.23}  & 2.00     & 3.38  & \textbf{1.99}  & 2.98             & \textbf{1.79}   & \textbf{2.97}  & 1.83          \\
\multicolumn{1}{|l}{} & webvideo                          & \textbf{29.77} & \textbf{31.51} & 33.94          & 36.29          & \textbf{27.65}   & \textbf{30.19}  & 32.18          & 34.53         \\
\multicolumn{1}{|l}{} & restaurant                        & 35.04 & 37.28 & \textbf{29.08} & \textbf{29.93} & \textbf{29.41}   & 31.29  & 29.62          & \textbf{30.90} \\
\multicolumn{1}{|l}{} & audiobooks                        & 1.66  & \textbf{0.31}  & \textbf{0.40}   & 0.50   & \textbf{0.00}       & \textbf{0.00}               & \textbf{0.00}              & \textbf{0.00}             \\
\multicolumn{1}{|l}{} & clinical                          & \textbf{5.51}  & \textbf{3.35}  & 7.33           & \textbf{4.34}  & \textbf{5.41}    & \textbf{3.18}   & 6.14           & 3.63          \\
\multicolumn{1}{|l}{} & socio-field                       & 8.22  & \textbf{3.86}  & \textbf{7.01}  & 5.18  & 7.05    & \textbf{3.49}    & \textbf{6.90}   & 3.91          \\
\hline
\multicolumn{1}{|l}{} & Mix-Headset                       & 14.93 & 17.33 & \textbf{14.81}          & \textbf{17.29}          & 14.92            & \textbf{17.30}            & \textbf{14.76}          & 17.32         \\
\multicolumn{1}{|c}{\small AMI}         & Mix-Lapel                         & 15.23 & 17.87 & \textbf{15.19} & \textbf{17.83}          & 15.18            & 17.94           & \textbf{15.13}          & \textbf{17.86}         \\
\multicolumn{1}{|l}{} & Array1-01                         & 15.62 & 18.98 & \textbf{15.51}          & \textbf{18.94}          & 15.48            & 18.29           & \textbf{15.35}          & \textbf{18.18}         \\
\hline
\end{tabular}
}
\vspace{-0.1cm}
\end{table}

\textbf{On the choice of kernels:} In Table \ref{tab:kernels}, we summarize a comparison of DERs achieved by MK-SGC-SC over a wide range of choices of kernels. The two best average ranks were achieved when four polynomials (\texttt{poly1-4}) were used along with the arccosine kernel of degree 1 (\texttt{arccos1}), or along with degree 0 (\texttt{arccos0-1}). Both choices achieved statistically comparable performances, as determined by the Wilcoxon signed-rank test~\cite{wilcoxon} for \texttt{poly1-4 + arccos1} (Test 1), or for \texttt{poly1-4 + arccos0-1} (Test 2). This justifies our choice of (\texttt{poly1-4 + arccos1}). However, including the cosine kernel, or exponential kernels (\texttt{exp1/2/3}), or considering all kernels generally performed worse. This implies adding more kernels does not help, a search over combinatorial kernel options may yield more improvements. In general, considering only one kernel was not as helpful either.

\begin{table}[!htb]
\centering
\vspace{-0.4cm}
\renewcommand{\arraystretch}{1.1}
\caption{DERs with MK-SGC-SC using different sets of kernels. $H_0$ and $H_1$ denote the null and alternative hypotheses, respectively.}
\label{tab:kernels}
\adjustbox{max width=\columnwidth}{
\begin{tabular}{lccccc}
\hline
Kernel       & \texttt{cosine}                         & \texttt{poly1}                           & \texttt{poly2}                          & \texttt{poly3}                          & \texttt{poly4}                          \\
\hline
Avg. Rank    & \multicolumn{1}{c}{8.73} & \multicolumn{1}{c}{9.53}  & \multicolumn{1}{c}{8.73} & \multicolumn{1}{c}{9.53} & \multicolumn{1}{c}{8.73} \\
Hyp. Test 1  & $H_1$                             & $H_1$                              & $H_1$                             & $H_1$                             & $H_1$                             \\
p-val Test 1 & \multicolumn{1}{c}{1.86E-09}   & \multicolumn{1}{c}{1.86E-09}    & \multicolumn{1}{c}{1.86E-09}   & \multicolumn{1}{c}{1.86E-09}   & \multicolumn{1}{c}{1.86E-09}   \\
Hyp. Test 2  & $H_1$                             & $H_1$                              & $H_1$                             & $H_1$                             & $H_1$                             \\
p-val Test 2 & \multicolumn{1}{c}{3.73E-09}   & \multicolumn{1}{c}{1.86E-09}    & \multicolumn{1}{c}{1.73E-06}   & \multicolumn{1}{c}{1.86E-09}   & \multicolumn{1}{c}{1.73E-06}   \\
\hline
\hline
Kernel       & \texttt{exp1}                           & \texttt{exp2}                            & \texttt{exp3}                           & \texttt{arccos0}                        & \texttt{arccos1}                        \\
\hline
Avg. Rank    & \multicolumn{1}{c}{13.20}       & \multicolumn{1}{c}{13.33} & \multicolumn{1}{c}{13.90}       & \multicolumn{1}{c}{7.80}        & \multicolumn{1}{c}{8.40}        \\
Hyp. Test 1  & $H_1$                             & $H_1$                              & $H_1$                             & $H_1$                             & $H_1$                             \\
p-val Test 1 & \multicolumn{1}{c}{1.86E-09}   & \multicolumn{1}{c}{1.86E-09}    & \multicolumn{1}{c}{1.86E-09}   & \multicolumn{1}{c}{1.86E-09}   & \multicolumn{1}{c}{1.86E-09}  \\
Hyp. Test 2  & $H_1$                             & $H_1$                              & $H_1$                             & $H_1$                             & $H_1$                             \\
p-val Test 2 & \multicolumn{1}{c}{1.86E-09}   & \multicolumn{1}{c}{1.86E-09}    & \multicolumn{1}{c}{1.86E-09}   & \multicolumn{1}{c}{3.73E-09}   & \multicolumn{1}{c}{1.86E-09}   \\
\hline
\hline
\makecell{Kernel \\(Consensus)}       & \texttt{poly1-4}                        & \makecell{\texttt{poly1-4} \\ +\texttt{arccos1}}                 & \makecell{\texttt{poly1-4} \\ +\texttt{arccos0-1}}              & \makecell{\texttt{poly1-4} \\ +\texttt{arccos0-1} \\ +\texttt{cosine}}       & \texttt{all}          \\
\hline
Avg. Rank    & \multicolumn{1}{c}{2.87} & \multicolumn{1}{c}{\textbf{2.50}}         & \multicolumn{1}{c}{\textbf{2.30}}        & \multicolumn{1}{c}{2.57} & \multicolumn{1}{c}{5.13} \\
Hyp. Test 1  & \textbf{\emph{H${}_\mathbf{0}$}}                             & \textbf{\emph{H${}_\mathbf{0}$}}                              & -                              & \textbf{\emph{H${}_\mathbf{0}$}}                              & $H_1$                             \\
p-val Test 1 & \multicolumn{1}{c}{\textbf{3.93E-01}}   & \multicolumn{1}{c}{\textbf{7.01E-01}}    & -                              & \multicolumn{1}{c}{\textbf{7.27E-02}}   & \multicolumn{1}{c}{1.30E-07}   \\
Hyp. Test 2  & \textbf{\emph{H${}_\mathbf{0}$}}                             & -                               & \textbf{\emph{H${}_\mathbf{0}$}}                              & \textbf{\emph{H${}_\mathbf{0}$}}                             & $H_1$                            \\
p-val Test 2 & \multicolumn{1}{c}{\textbf{1.85E-01}}   & \multicolumn{1}{c}{-}    & \multicolumn{1}{c}{\textbf{7.01E-01}}   & \multicolumn{1}{c}{\textbf{4.82E-01}}   & \multicolumn{1}{c}{3.51E-06} \\
\hline
\end{tabular}
}
\end{table}
\vspace{-.2cm}
\textbf{Ablation Studies:} We conduct several ablation studies to verify the choices made for MK-SGC-SC, and to consider alternatives. First we examine the necessity of considering graph sparsification instead of only operating on the fused multiple kernel similarities. The `w/o sparsity' column of results in Table \ref{tab:ablation} shows significant increase in DERs achieved compared to the baseline DERs achieved by MK-SGC-SC (shown in the first column of results), clearly denoting the necessity of a graph sparsification procedure. Next, we consider the use of a normalized symmetric Laplacian $L=D^{-1/2}AD^{-1/2}$ instead of the unnormalized Laplacian $L=D-A$. The `norm-$L$' column of results in Table \ref{tab:ablation} generally shows increase in DERs achieved, with the only exceptions in the cases of court, meeting, and restaurant. Finally, we consider possible weighting schemes $K^*=\sum_l w_lK^l$ during fusion of the kernel similarities. We generally observed no significant effect of weighting schemes, an example of which is provided in the `entropy-$w$' column of results in Table \ref{tab:ablation}, where entropy of the kernel similarities was measured and converted to weights.

\begin{table}[!htb]
\centering
\caption{DERs (i) without graph sparsification, (ii) with normalized Laplacian, (iii) with entropy-based kernel weighting.}
\label{tab:ablation}
\adjustbox{max width=\columnwidth}{
\begin{tabular}{ll|cc|cc|cc|cc|}
\cline{3-10}
& \multicolumn{1}{l}{}              & \multicolumn{2}{|c}{MK-SGC-SC}                      & \multicolumn{2}{|c}{w/o sparsity}           & \multicolumn{2}{|c}{norm-$L$}      & \multicolumn{2}{|c|}{entropy-$w$}         \\
& \multicolumn{1}{l}{}              & \multicolumn{1}{|l}{Dev} & \multicolumn{1}{l}{Eval} & \multicolumn{1}{|l}{Dev} & \multicolumn{1}{l}{Eval} & \multicolumn{1}{|l}{Dev} & \multicolumn{1}{l}{Eval} & \multicolumn{1}{|l}{Dev} & \multicolumn{1}{l|}{Eval} \\
\hline
\multicolumn{1}{|l}{} & broadcast & 1.64                    & 2.70                      & 13.48                   & 9.05                     & 6.51                    & 3.61                     & 1.64                    & 2.70                      \\
\multicolumn{1}{|l}{} & court     & 4.64                    & 2.79                     & 34.12                   & 35.12                    & 2.63           & 2.43            & 4.64                    & 2.79                     \\
\multicolumn{1}{|l}{} & cts                               & 8.20                     & 6.62            & 33.2                    & 29.88                    & 8.44                    & 6.67                     & 8.20                     & 6.63                     \\
\multicolumn{1}{|l}{} & maptask                           & 1.88                    & 0.94                     & 13.56                   & 5.73                     & 3.06                    & 1.44                     & 1.88                    & 0.94                     \\
\multicolumn{1}{|l}{} & meeting                           & 16.29                   & 19.09                    & 38.55                   & 46.6                     & 16.67                   & 15.81           & 16.29                   & 19.09                    \\
\multicolumn{1}{|c}{\small DIHARD-III}  & socio-lab & 3.23           & 2.00                        & 16.64                   & 7.00                        & 5.88                    & 5.56                     & 3.27                    & 2.00                        \\
\multicolumn{1}{|l}{} & webvideo                          & 29.77          & 31.51                    & 36.08                   & 43.79                    & 36.42                   & 36.35                    & 29.78                   & 31.44           \\
\multicolumn{1}{|l}{} & restaurant                        & 35.04          & 37.28           & 55.11                   & 64.11                    & 30.68          & 31.7            & 35.07                   & 38.49                    \\
\multicolumn{1}{|l}{} & audiobooks                        & 1.66                    & 0.31                     & 0.11                    & 0.20                      & 31.85                   & 33.03                    & 1.66                    & 0.31                     \\
\multicolumn{1}{|l}{} & clinical                          & 5.51         & 3.35                     & 20.23                   & 27.63                    & 7.38                    & 5.50                      & 5.54                    & 3.35                     \\
\multicolumn{1}{|l}{} & socio-field                       & 8.22                    & 3.86                     & 25.11                   & 20.10                     & 9.33                    & 9.06                     & 8.11           & 3.86                     \\
\hline
\end{tabular}
}
\end{table}

We also analyzed the choice for the number of neighbors $c=15$, by running MK-SGC-SC for values of $c$ ranging from 11 to 19, in increments of two. On each data split the DER across all $c$ was noted, following which the deviations from the minimum DER was measured for all $c$. Table \ref{tab:tab_neighbors} summarizes the deviations using average ranks and Wilcoxon signed-rank test, where we observe that the deviations were generally the smallest for $c=15$, however the deviations for other $c$ were also comparable, indicating that MK-SGC-SC behaves robustly around $c=15$ neighbors.

\begin{table}[!htb]
\centering
\vspace{-0.4cm}
\caption{Deviation from the minimum DER achieved by MK-SGC-SC, when considering different number of neighbors $c$.}
\label{tab:tab_neighbors}
\adjustbox{max width=0.8\columnwidth}{
\begin{tabular}{lccccc}
\hline
          & c=11                           & c=13                           & c=15                           & c=17                           & c=19                           \\
          \hline
Avg. Rank & \multicolumn{1}{c}{3.13} & \multicolumn{1}{c}{2.70}        & \multicolumn{1}{c}{\textbf{2.27}} & \multicolumn{1}{c}{2.43} & \multicolumn{1}{c}{2.83} \\
Hyp. Test & $H_0$                             & $H_0$                             & -                              & $H_0$                             & $H_1$                             \\
p-val     & \multicolumn{1}{c}{0.22} & \multicolumn{1}{c}{0.63} & -                              & \multicolumn{1}{c}{0.32} & \multicolumn{1}{c}{0.03} \\
\hline
\end{tabular}
}
\vspace{-0.1cm}
\end{table}

\vspace{-0.25cm}
\section{Conclusion}
\label{sec:conclusion}
\vspace{-0.25cm}
In the context of spectral clustering based methods for unsupervised speaker diarization, our findings can be summarized as - \emph{measuring multiple kernel similarities of speaker embeddings makes the construction of cluster-aware sparse graphs easier.} This is based on the proposed MK-SGC-SC, which measures four polynomial kernels and a degree one arccosine kernel on speaker embeddings, to thereafter construct a sparse graph in a principled manner, on which spectral clustering identifies the clusters. The excelling unsupervised diarization performance of MK-SGC-SC across multiple speech corpora indicates that it is a reliable and generalizable approach, while having the same time complexity of standard spectral clustering, and the space complexity is higher only in a multiplicative factor of the number of kernels considered. This work raises questions for further research: how to explore kernel–embedding combinations, optimize sparse graphs, and narrow the DER gap in overlapped speech.

\section*{Acknowledgment}
The primary author would like to thank his doctoral advisory committee members at RKMVERI for their continued feedback and guidance. He also expresses sincere gratitude to the Linguistic Data Consortium (LDC) for awarding the LDC Data Scholarship, which enabled access to the DIHARD-III dataset.
\bibliographystyle{IEEEbib}
\bibliography{refs}

\end{document}